# Ein neuer Ansatz zur Frequenzmodellierung im Versicherungswesen

Dietmar Pfeifer
Institut für Mathematik
Fakultät V
Carl-von-Ossietzky Universität Oldenburg

Stand: 04.08.2023 (corrected version)

**Zusammenfassung** Im kollektiven Modell der Risikotheorie wird üblicherweise zwischen der Schadenfrequenz bzw. deren Verteilung und der Einzelschadenhöhe bzw. deren Verteilung unterschieden. Für die Schadenfrequenzverteilung werden meist „klassische" Verteilungen wie Binomial-, Negative Binomial- oder Poissonverteilung unterstellt, die aber gelegentlich nicht passend sind. In dieser Arbeit wählen wir für die Schadenfrequenz einen anderen Ansatz über einen zufälligen prozentualen Anteil an der Zahl der Versicherungsverträge (sog. Betroffenheitsgrad). Dieser Ansatz erlaubt u.a. eine statistische Überprüfung durch Q-Q-Plots und lässt sich auch leicht auf Schadenhöhenverteilungen übertragen.

**A new approach to claims frequency modelling in insurance**

**Abstract.** The collective risk model differentiates usually between claims frequencies (and their distribution) and claim sizes (and their distribution). For the claims frequencies typically "classical" discrete distributions are considered, such as Binomial-, Negative binomial- or Poisson distributions. Since these distributions sometimes do not really fit to the data we propose a different approach here for claim frequencies via random proportions of the number of insurance contracts. This approach also allows for a statistical goodness-of-fit test via quantile-quantile-plots and can likewise be applied to the modelling of claim size distributions.

## 1. Vorbemerkung

Das kollektive Modell der Risikotheorie, welches üblicherweise auch zur Tarifierung in der Sachversicherung verwendet wird, zeichnet sich durch die folgenden beiden Komponenten aus:

1. Die Frequenzverteilung, die die Anzahl der durch Schäden betroffenen Verträge statistisch beschreibt
2. Die Schadenhöhenverteilung, die den individuell eingetretenen Schaden statistisch beschreibt.

Gängige Verteilungen für die Frequenz sind die Binomial-, die Poisson- und die negative Binomialverteilung, vgl. Heilmann und Schröter (2014), Kap. 2 und 3.

---

D. Pfeifer (✉)
Institut für Mathematik, Schwerpunkt Versicherungs- und Finanzmathematik, Carl von Ossietzky Universität Oldenburg, Oldenburg, Deutschland
E-Mail: dietmar.pfeifer@uni-oldenburg.de

In der Praxis haben sich diese Verteilungen allerdings auf Grund empirischer Analysen oft als nicht brauchbar erwiesen. Daher soll in dieser Arbeit ein anderer, mit der Praxis besser in Einklang zu bringender Ansatz für die Modellierung der Frequenz vorgestellt werden, der sich in ähnlicher Weise auch auf die Modellierung von Einzelschadenhöhen übertragen lässt, vor allem, wenn die Versicherungsverträge durch unterschiedlich hohe Versicherungssummen oder maximale Entschädigungssummen charakterisiert sind.

Die Methodik wird beispielhaft anhand einer Versicherungssparte mit 4 Tarifen illustriert.

## 2. Einführung

Die zentrale Idee ist hier die Beschreibung der Frequenz durch einen zufälligen Anteil $\xi$ der von Schäden betroffenen Verträge, der jährlich variieren kann und den wir Betroffenheitsgrad nennen wollen. Da $\xi$ zwischen 0 und 1 liegt, bietet sich natürlicher Weise eine Modellierung durch eine Betaverteilung an. Die Dichtefunktion $f_B$ einer Betaverteilung ist gegeben durch

$$f_B(x) = \frac{x^{\alpha-1}(1-x)^{\beta-1}}{\text{Be}(\alpha,\beta)} \text{ für } 0 < x < 1 \text{ mit den Parametern } \alpha, \beta > 0. \tag{1}$$

Hierbei bezeichnet Be die Euler'sche Betafunktion, die durch $\text{Be}(\alpha,\beta) = \frac{\Gamma(\alpha)\Gamma(\beta)}{\Gamma(\alpha+\beta)}$ mit der Euler'schen Gamma-Funktion $\Gamma(x) = \int_0^\infty t^{x-1} e^{-t}\, dt$ für $x > 0$ definiert ist.

Ein Betaverteiltes Risiko $\xi$ besitzt die folgenden charakteristischen Kenngrößen:

$$\text{Erwartungswert } E(\xi) = \frac{\alpha}{\alpha+\beta} \text{ und Varianz } Var(\xi) = \frac{\alpha\beta}{(\alpha+\beta)^2(\alpha+\beta+1)}. \tag{2}$$

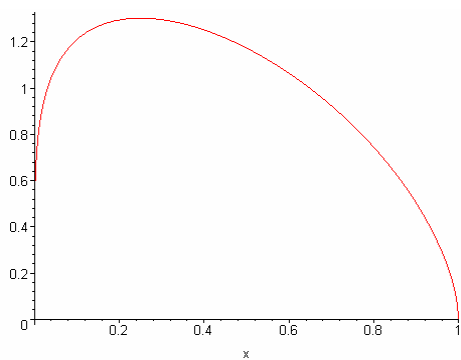
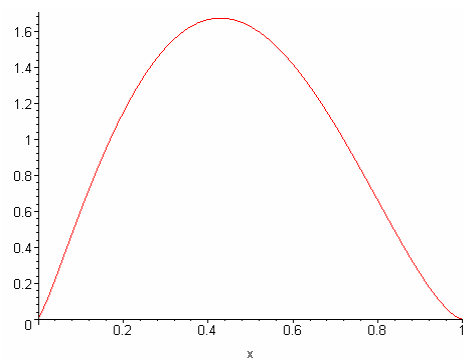

$\alpha = 1,2 \quad \beta = 1,6$ $\qquad\qquad\qquad$ $\alpha = 2,2 \quad \beta = 2,6$

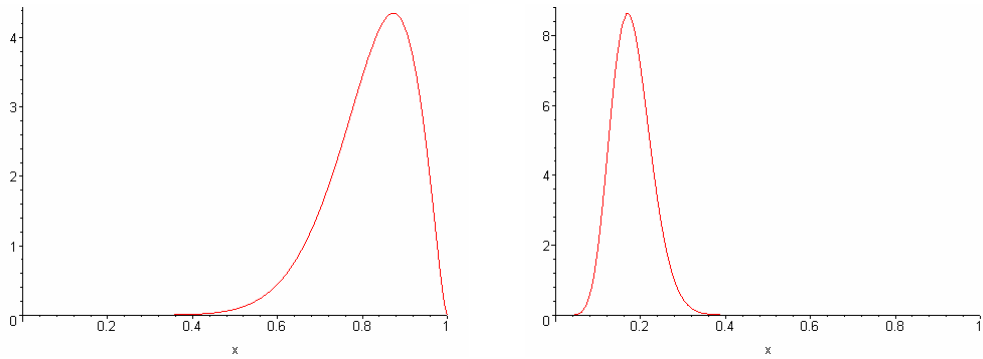

$\alpha = 12 \quad \beta = 2,6$ $\qquad\qquad\qquad$ $\alpha = 12 \quad \beta = 55$

graphische Darstellung von Dichten $f_B$ verschiedener Betaverteilungen

Eine Schätzung der Parameter $\alpha$ und $\beta$ der Verteilung von $\xi$ kann mit Hilfe der *Momentenmethode* (vgl. Czado und Schmidt (2011), Abschnitt 3.1.2) durch Gleichsetzung von empirischen und theoretischen Momenten aus Beobachtungen $p_1,\cdots,p_n$ von $\xi_1,\cdots,\xi_n$ (alle mit derselben Betaverteilung) wie folgt durchgeführt werden:

$$\hat{\alpha} = \frac{\mu_b\left(\mu_b - \mu_b^2 - \sigma_b^2\right)}{\sigma_b^2} \quad \text{und} \quad \hat{\beta} = \frac{1-\mu_b}{\mu_b}\hat{\alpha} \qquad (3)$$

Hierbei bezeichnet $\mu_b = \dfrac{1}{n}\sum_{k=1}^{n} p_k$ den Mittelwert und $\sigma_b^2 = \dfrac{1}{n-1}\sum_{k=1}^{n}(p_k - \mu_b)^2$ die empirische Varianz der Beobachtungen.

Für das Testen der Hypothese, dass die Beobachtungen tatsächlich einer Betaverteilung (zumindest näherungsweise) genügen, bietet sich ein Verfahren ähnlich wie in Pfeifer (2019) an. Dazu trägt man in einem Quantil-Quantil-Plot die der Größe nach angeordneten beobachteten Beobachtungen gegen die Quantile $q_k$ der geschätzten Betaverteilung ab, die über die geschätzten Parameter wie folgt definiert sind:

$$q_k = Q\left(\frac{k}{n+1}; \hat{\alpha}, \hat{\beta}\right) \text{ für } k = 1,\cdots,n. \qquad (4)$$

Dabei bezeichnet $Q(\bullet;\alpha,\beta)$ die zur Betaverteilung mit den Parametern $\alpha,\beta > 0$ gehörige Quantilfunktion (d.h. die inverse Verteilungsfunktion). Leider lässt sich das in Pfeifer (2019) beschriebene Verfahren nicht direkt anwenden, weil die Betaverteilungsfamilie keine Lagen-Skalen-Familie darstellt und auch nicht durch einfache Transformationen darauf zurückgeführt werden kann. Trotzdem kann die dort eingeführte Teststatistik $T_n = -\ln(1-\rho_n)$ auch hier verwendet werden, wobei wieder $\rho_n$ den empirischen Korrelationskoeffizienten zwischen den Quantilen $q_k$ und den der Größe nach angeordneten Beobachtungen bezeichnet. Allerdings lässt sich die Verteilung von $T_n$ nicht ohne weiteres durch eine Gamma- oder Normalverteilung beschreiben, wie in Pfeifer (2019). Für einen geeigneten Signifikanztest kann man aber einen Monte-Carlo-Test verwenden, indem man $n$ Betaverteilte Daten simulativ erzeugt und für den Test die empirische Verteilung der Testgröße verwendet. Monte-Carlo-Tests werden genauer in Zhu (2005) beschrieben.

In der Praxis sind die jährlichen Betroffenheitsgrade aber nicht so ohne weiteres bestimmbar. Sie können nur indirekt über das Verhältnis der jährlich von Schäden betroffenen Verträge zu der Gesamtzahl der jährlichen Verträge im Portfolio geschätzt werden, wobei sich die Bestände in der Regel von Jahr zu Jahr verändern. Dabei wird unterstellt, dass die jährlichen Betroffenheitsgrade $\xi_1, \cdots, \xi_n$ derselben Betaverteilung folgen, deren Parameter in gewisser Weise charakteristisch für die betrachtete Vertragsart sind.

Bezeichnet man mit $M_1, \cdots, M_n$ die jährlichen Vertragszahlen und mit $A_1, \cdots, A_n$ die Zahl der jährlich von Schäden betroffenen Verträge, so ist

$$\hat{p}_i = \frac{A_i}{M_i}, \ i = 1, \cdots, n \tag{5}$$

ein Schätzwert für den realisierten Betroffenheitsgrad $p_i$ von $\xi_i$ im Jahr $i$. Wenn man sich vorstellt, dass nach Realisation eines Betaverteilten Betroffenheitsgrades $p_i$ aus den $M_i$ im Jahr $i$ vorhandenen Verträgen die mit Schäden belasteten Verträge „zufällig" mit einem Null-Eins-Experiment mit „Trefferwahrscheinlichkeit" $p_i$ ausgewählt werden, so entspricht die (bedingte) Verteilung der Anzahl $A_i$ bei gegebenem $p_i$ einer klassischen Binomialverteilung, gegeben durch

$$P(A_i = k | \xi_i = p_i) = \binom{M_i}{k} p_i^k (1-p_i)^{M_i - k}, \ k = 0, \cdots, M_i \tag{6}$$

und den bedingten Momenten Erwartungswert und Varianz

$$E(A_i | \xi_i = p_i) = M_i p_i \text{ und } Var(A_i | \xi_i = p_i) = M_i p_i (1-p_i), \ k = 0, \cdots, M_i. \tag{7}$$

Die unbedingte Verteilung von $A_i$ ist eine Mischverteilung, die sich aus Gewichtung der in (6) gegebenen Wahrscheinlichkeiten mit den durch die Betroffenheitsgrade gegebenen Wahrscheinlichkeiten ergibt, in Formeln:

$$P(A_i = k) = \int_0^1 P(A_i = k | \xi_i = p_i) f_B(p_i) dp_i = \binom{M_i}{k} p_i \int_0^1 \frac{p_i^{k+\alpha-1}(1-p_i)^{M_i-k+\beta-1}}{\text{Be}(\alpha,\beta)} dp_i$$

$$= \binom{M_i}{k} \frac{\text{Be}(k+\alpha, M_i-k+\beta)}{\text{Be}(\alpha,\beta)} = \binom{M_i}{k} \frac{\Gamma(k+\alpha)\Gamma(M_i-k+\beta)}{\Gamma(\alpha)\Gamma(\beta)} \frac{\Gamma(\alpha+\beta)}{\Gamma(M_i+\alpha+\beta)} \quad (8)$$

für $k = 0, \cdots, M_i$. Diese Art Verteilung wird in der Literatur als *Beta-Binomialverteilung* bezeichnet, vgl. Johnson et al. (2005), Kapitel 6.2.2. Für die unbedingten Momente Erwartungswert und Varianz ergibt sich entsprechend

$$E(A_i) = E(E(A_i | \xi_i)) = \int_0^1 E(A_i | \xi_i = p_i) f_B(p_i) dp_i$$

$$= M_i \int_0^1 p_i f_B(p_i) dp_i = M_i \cdot E(\xi_i) = M_i \frac{\alpha}{\alpha+\beta} \quad \text{und} \quad (9)$$

$$Var(A_i) = E(Var(A_i | \xi_i)) + Var(E(A_i | \xi_i)) = M_i \int_0^1 p_i(1-p_i) f_B(p_i) dp_i + M_i^2 Var(\xi_i)$$

$$= M_i \frac{\Gamma(1+\alpha)\Gamma(1+\beta)}{\Gamma(\alpha)\Gamma(\beta)} \frac{\Gamma(\alpha+\beta)}{\Gamma(2+\alpha+\beta)} + M_i^2 \frac{\alpha\beta}{(\alpha+\beta)^2(\alpha+\beta+1)}$$

$$= M_i \frac{\alpha\beta(\alpha+\beta+M_i)}{(\alpha+\beta)^2(\alpha+\beta+1)} = M_i^2 \cdot Var(\xi_i) \cdot \left(1 + \frac{\alpha+\beta}{M_i}\right) \quad (10)$$

nach (2) und (8) (setze dort $M_i = 1$ und $k = 1$), vgl. die Beziehung (6.12) in Johnson et al. (2005) und Becker et al. (2014), Beziehung (10.7) für die Varianzformel. Für den empirischen Betroffenheitsgrad $\hat{p}_i = \frac{A_i}{M_i}$ erhält man hieraus

$$E(\hat{p}_i) = E(\xi_i) = \frac{\alpha}{\alpha+\beta} \quad \text{und} \quad Var(\hat{p}_i) = \frac{1}{M_i^2} Var(A_i) = \left(1 + \frac{\alpha+\beta}{M_i}\right) \cdot Var(\xi_i). \quad (11)$$

Für große Werte von $M_i$ wird $Var(\hat{p}_i)$ näherungsweise gleich $Var(\xi_i)$. Ferner ist

$$E(\hat{p}_i | \xi_i = p_i) = p_i \quad \text{und} \quad Var(\hat{p}_i | \xi_i = p_i) = \frac{1}{M_i} p_i(1-p_i) \leq \frac{1}{4M_i}, \quad (12)$$

d.h. bei großem $M_i$ weicht der jährliche empirische Betroffenheitsgrad $\hat{p}_i = \frac{A_i}{M_i}$ nur in sehr geringem Maße von dem für das betreffende Jahr realisierten Betroffenheitsgrad $p_i$ ab. In der Praxis kann man deshalb die empirische Betroffenheitsgrade $\hat{p}_i = \frac{A_i}{M_i}$ mit den Betaverteilten realisierten Betroffenheitsgraden $p_i$ „gleichsetzen" und somit für eine statistische Analyse wie in (3) heranziehen. Dadurch lässt sich die Hypothese einer Betaverteilung für die Betroffenheitsgrade bzw. äquivalent die Hypothese des Vorliegens einer Beta-Binomialverteilung für die Schadenzahlen testen.

Wir veranschaulichen das Verfahren an Frequenzzahlen aus einer Versicherungssparte mit 4 Tarifen aus fünf Beobachtungsjahren.

| Jahr | Tarif A Verträge | betroffen | Tarif B Verträge | betroffen | Tarif C Verträge | betroffen | Tarif D Verträge | betroffen |
|---|---|---|---|---|---|---|---|---|
| 2011 | 8.805 | 327 | 4.276 | 149 | 1.094 | 42 | 21.984 | 695 |
| 2012 | 12.754 | 523 | 3.387 | 131 | 836 | 23 | 24.250 | 870 |
| 2013 | 16.185 | 644 | 2.723 | 75 | 656 | 26 | 26.378 | 921 |
| 2014 | 20.675 | 831 | 2.177 | 71 | 523 | 13 | 29.306 | 1.102 |
| 2015 | 26.567 | 1.009 | 1.767 | 44 | 435 | 9 | 33.751 | 1.192 |

Tab. 1

Hieraus ergeben sich folgende empirische Betroffenheitsgrade $\hat{p}_i$ in % und geschätzte Parameter (gerundet):

| Jahr | Tarif A | Tarif B | Tarif C | Tarif D |
|---|---|---|---|---|
| 2011 | 3,71% | 3,48% | 3,84% | 3,16% |
| 2012 | 4,10% | 3,87% | 2,75% | 3,59% |
| 2013 | 3,98% | 2,75% | 3,96% | 3,49% |
| 2014 | 4,02% | 3,26% | 2,49% | 3,76% |
| 2015 | 3,80% | 2,49% | 2,07% | 3,53% |
| $\mu_b$ | 3,92% | 3,17% | 3,02% | 3,51% |
| $\sigma_b$ | 0,16% | 0,55% | 0,84% | 0,22% |
| $\hat{\alpha}$ | 572 | 32 | 13 | 249 |
| $\hat{\beta}$ | 14.007 | 967 | 402 | 6.838 |

Tab. 2

Die folgende Graphik zeigt die zugehörigen geschätzten Beta-Dichten für die Betroffenheitsgrade:

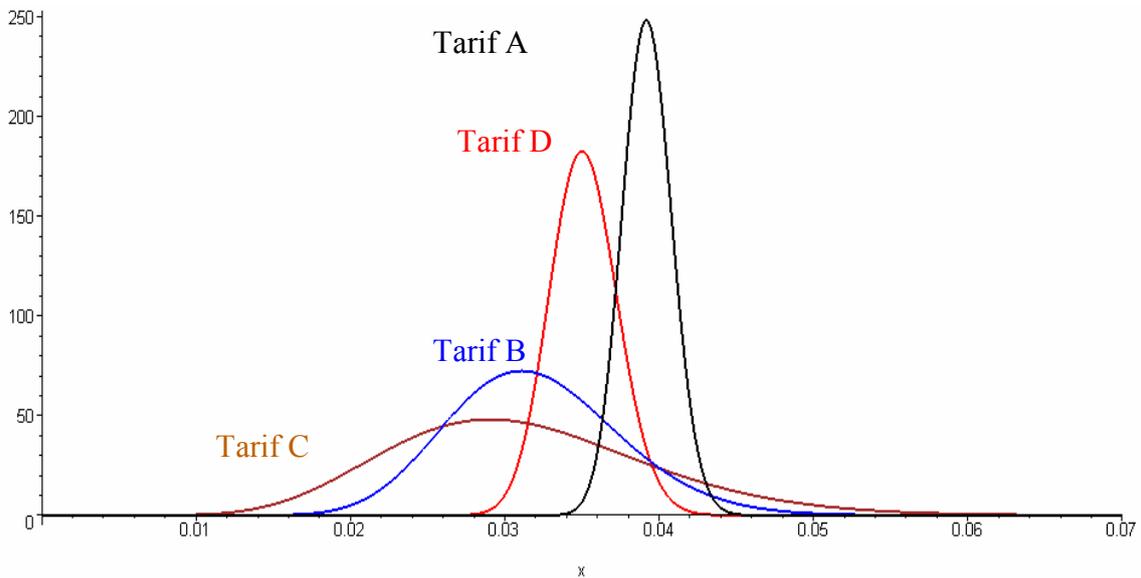

Es zeigen sich deutliche Unterschiede in der Lage und in der Streuung der Betroffenheitsgrade für die verschiedenen Tarife.

Die folgenden Graphiken zeigen die zugehörigen Quantil-Quantil-Plots zusammen mit den Testgrößen $T_n$ und den *p*-Werten aus dem Monte-Carlo-Test:

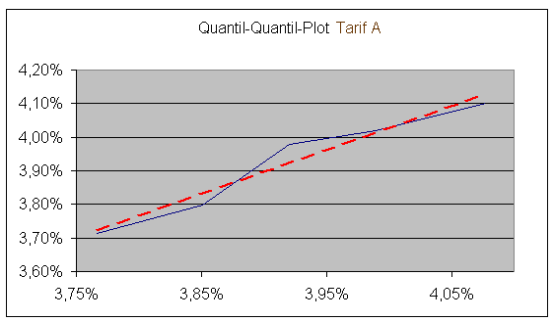

$T_n = 3,663 \quad p\text{-Wert} = 71,77\%$

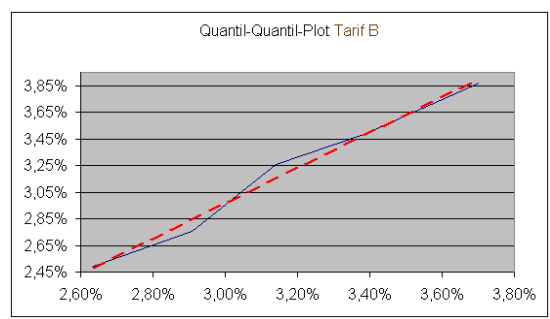

$T_n = 4,748 \quad p\text{-Wert} = 94,22\%$

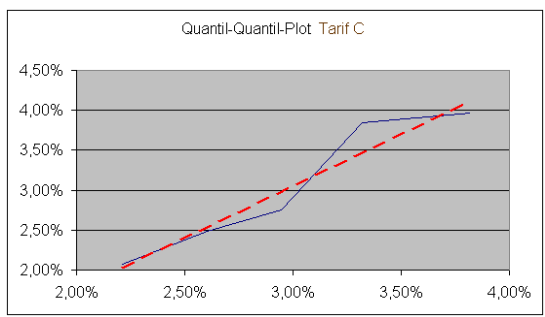

$T_n = 3,219 \quad p\text{-Wert} = 42,63\%$

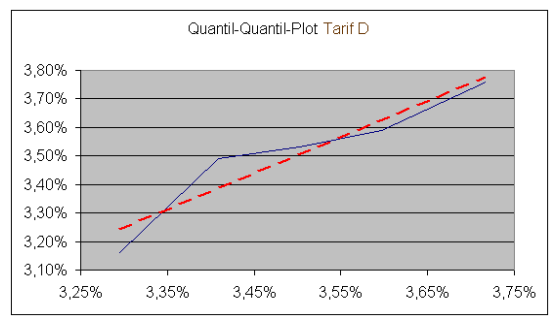

$T_n = 2,924 \quad p\text{-Wert} = 67,98\%$

Die Analyse zeigt, dass für alle vier Tarife die Annahme einer Betaverteilung für die Betroffenheitsgrade akzeptabel ist.

## 3. Literatur